\documentclass{emulateapj}

\usepackage{graphicx}
\usepackage{amsmath}
\usepackage{natbib}
\usepackage{color}
\usepackage{times}

\shorttitle{The final spin from the coalescence of aligned-spin black-hole
binaries}
\shortauthors{Rezzolla et al.}

\begin{document}

\title{The final spin from the coalescence of aligned-spin black-hole
binaries}

\author{Luciano Rezzolla,\altaffilmark{1,3} 
   Peter Diener,\altaffilmark{2,3} 
   Ernst Nils Dorband,\altaffilmark{1} 
   Denis Pollney,\altaffilmark{1} 
   Christian Reisswig,\altaffilmark{1}
   Erik Schnetter,\altaffilmark{2,3} 
   Jennifer Seiler,\altaffilmark{1}}
\altaffiltext{1}{Max-Planck-Institut f\"ur Gravitationsphysik,
Albert-Einstein-Institut, Potsdam-Golm, Germany}

\altaffiltext{2}{Center for Computation \& Technology, Louisiana State
University, Baton Rouge, LA, USA}

\altaffiltext{3}{Department of Physics and Astronomy, Louisiana State
University, Baton Rouge, LA, USA }

\begin{abstract}
	Determining the final spin of a black-hole (BH) binary is a
	question of key importance in astrophysics
	. Modelling this quantity in general is made difficult
	by the fact that it depends on the 7-dimensional space of
	parameters characterizing the two initial black
	holes. However, in special cases, when symmetries can be
	exploited, the description can become 
	simpler.
	For black-hole binaries with unequal masses but with equal
	spins which are aligned with the orbital angular momentum, we
	show that the use of recent simulations and basic but exact
	constraints derived from the extreme mass-ratio limit allow to
	model this quantity with a simple analytic expression. Despite
	the simple dependence, the expression models very
	accurately all of the available estimates, with errors of a
	couple of percent at most. We also discuss how to use the
	fit 
	to predict when a Schwarzschild BH is
	produced by the merger of two spinning BHs, when the
	total angular momentum of the spacetime ``flips'' sign, or
	under what conditions the final BH is ``spun-up'' by
	the merger. 
	Finally, suggest an extension of the
	fit 
	to include unequal-spin binaries, thus
	potentially providing a \textit{complete} description of the
	final spin from the coalescence of generic black-hole binaries
	with spins aligned to the orbital angular momentum.
\end{abstract}

\keywords{black hole physics -- relativity -- gravitational waves --
stars: statistics}

\maketitle

\section{INTRODUCTION}
\label{intro}

The determination of the final spin of a BH binary is a
question of key importance in astrophysics. Modelling this in general
is made difficult by the fact that it depends on the 7-dimensional
space of parameters characterizing the two initial BHs. However, in
special cases, when symmetries can be exploited, the description can
be much simpler.

Several recent studies have shed light on the remnant of the merger
process. Using conservation principles, Hughes and
Blandford~\citep{Hughes_Blandford:2003} argued that mergers
rarely lead to rapidly rotating objects. \citet{Gonzalez:2006md}
numerically evolved a sequence of non-spinning unequal-mass BHs,
arriving at detailed estimates of the radiated energy and angular
momentum.  In a series of papers~\citep{Koppitz:2007ev,
Pollney:2007ss, Rezzolla_etal:2007a} we have studied the parameter
space of mergers of equal-mass BH binaries whose spins are aligned
with the orbital angular momentum but otherwise arbitrary.  The
findings agree well with independent numerical
evolutions~\citep{Campanelli_etal:2007,Herrmann:2007ac}, as well as
more recent studies of models with initial spins up to
$J/M^2=0.8$~\citep{Marronetti_etal:2007}. An important result of these
studies has been the determination of simple (quadratic) fitting
formulas for the recoil velocity and spin of the merger remnant as a
function of the initial BH parameters~\citep{Rezzolla_etal:2007a}.

A number of analytical approaches have been developed over the years
to determine the final spin of a binary coalescence~\citep{
Damour:2001tu, Buonanno_Damour:2000, Buonanno_etal:2006,
Damour_Nagar:2007a, Boyle:2007sz}. Very recently, an interesting
method, inspired by the dynamics of a test particle around a Kerr BH,
has been proposed for generic binaries~(\citet{Buonanno_etal:2007b},
BKL hereafter). The approach assumes that the angular momentum of the
final BH is the sum of the individual spins and of the orbital angular
momentum of a test particle on the last-stable orbit of a Kerr BH with
the same spin parameter as that of the final BH.

Here, we combine the data obtained in recent simulations to provide a
phenomenological but analytic estimate for the final spin in a binary
BH system with arbitrary mass ratio and spin ratio, but in which the
spins are constrained to be parallel to the orbital angular
momentum. Our numerical simulations have been carried out using the
CCATIE code~\citep{Pollney:2007ss}. In addition to the data presented
in \citet{Rezzolla_etal:2007a}, we add three simulations of
equal-mass, high-spin binaries and three simulations of unequal-mass,
spinning binaries (see Table~\ref{tableone}).  Other data is taken
from 
unequal-mass, nonspinning
binaries~\citep{Gonzalez:2006md,Berti_etal:2007,Buonanno_etal:2007a},
and of equal-mass, spinning
binaries~\citep{Rezzolla_etal:2007a,Marronetti_etal:2007}; all of the
AEI data is summarized in Table~\ref{tableone}. To avoid the possible
contamination from the errors associated with high-spin binaries
reported by~\citet{Marronetti_etal:2007}, we have not considered
binaries with initial spin $|J/M^2| \geq 0.75$ reported in the
literature~\citep{Campanelli_etal:2007,Marronetti_etal:2007}. We have,
however, considered estimates of high-spin binaries (\textit{cf.}, 
Table~\ref{tableone}), for which we know the spins remain essentially
constant prior to merger, with changes less than
$0.5\%$~\citep{Pollney:2007ss}, and that are very well captured by the
fit.

\section{Methods and Results} 

We start by considering the final spin $a_{\rm fin}$ as a function of
the two free variables in the problem: the symmetric mass ratio $\nu
\equiv M_1M_2/(M_1+M_2)^2$ and the spin of the initial
BHs $a \equiv J/M^2$, \textit{i.e.,} $a_{\rm fin} \equiv
{J_{\rm fin}}/{M^2_{\rm fin}} = a_{\rm fin} (a,\,\nu)$. (Note $a$ is
dimensionless and not the angular momentum per unit mass.) By
construction $a_1=a_2=a$, and ${\vec a}/|{\vec a}| = \pm {\vec
L}/|{\vec L}|$, where ${\vec L}$ is the orbital angular momentum. We
next express $a_{\rm fin}$ as a third-order polynomial of $\nu$ and
$a$
\begin{eqnarray}
a_{\rm fin} = &&s_0          + s_1 a + s_2 a^2 + s_3 a^3 + 
              s_4 a^2\nu  + s_5 a\nu^2 + \nonumber \\
              &&t_0 a \nu + t_1 \nu + t_2 \nu^2 + t_3 \nu^3\,.
\label{eq:1}
\end{eqnarray}
%
Expression~(\ref{eq:1}) is a lowest-order \textit{ansatz}. It intends
to capture the behaviour of a function known exactly only in the
extreme mass-ratio limit (EMRL) and which has support from numerical
simulations in two restricted regimes: \textit{i.e.}, $\nu=1/4;\ 0
\leq |a| \lesssim 0.75$ and $0.16 \lesssim \nu \leq 1/4;\
a=0$. A-priori there is no reason to believe 
expectation that $a_{\rm fin}(\nu,a)$ is
that the proposed fit will capture the
general behaviour well, but in fact it does.

\begin{table}
\caption{\label{tableone}Initial parameters of the new binaries
  computed at the AEI. The different columns contain the initial spin
  $a$, the symmetric mass ratio $\nu$, half of the initial separation
  $x/M=\frac{1}{2}(x_1-x_2)$, the dimensionless initial angular momentum
  $\tilde{J}=J/(\mu M)$, the numerical and fitted values for $a_{\rm
  fin}$ and the corresponding relative error.}
\vspace*{1cm}
\begin{center}
\begin{tabular}{lrcccccc}
~					&
\multicolumn{1}{c}{$a$}		&
{$\nu$} 				&
{$x/M$} 			        &
{$\tilde{J}$}                           &
{$a_{_{\mathrm{fin}}}$}	                &
{$a_{_{\mathrm{fin}}}^{\rm fit}$}	&
\multicolumn{1}{l}{\!\!\!\!$|$err.$|$\,(\%)}\\
\tableline
\hline
$t8$        & -0.5840 & 0.2500 & 3.1712 & 2.432 & 0.4955 & 0.4981 & 0.53 \\  
$ta8$       & -0.3000 & 0.2500 & 3.7078 & 3.000 & 0.5941 & 0.5927 & 0.23 \\  
$tb8$       & -0.8000 & 0.2500 & 3.8082 & 2.200 & 0.4224 & 0.4227 & 0.08 \\  
$tb8{\ell}$ & -0.8000 & 0.2500 & 4.8600 & 2.400 & 0.4266 & 0.4227 & 0.92 \\  
$p1$        & -0.8000 & 0.1580 & 3.2733 & 0.336 & 0.0050 & 0.0046 & 9.89 \\  
$p2$        & -0.5330 & 0.1875 & 3.3606 & 1.872 & 0.2778 & 0.2794 & 0.57 \\  
$p3$        & -0.2667 & 0.2222 & 3.4835 & 2.883 & 0.5228 & 0.5216 & 0.23 \\  
\hline
\end{tabular}
\end{center}
\end{table}

Given the available 
numerical estimates
, it is 
possible to calculate the coefficients $s_0$--$s_5$, and
$t_0$--$t_3$ by simply performing a two-dimensional (2D) least-square
fit of the data. This, however, would require a lot of care and is
likely to lead to inaccurate estimates
. This is
because 
the space of parameters presently
accessible to numerical simulations is rather small. Reliable results
are in fact available only for spins $|a| \lesssim 0.8$ and mass
ratios $q \equiv M_2/M_1 \gtrsim 0.25$ and thus corresponding to $\nu
\gtrsim 0.16$. However, it is possible to exploit \textit{exact}
results which hold in the EMRL, \textit{i.e.}, for $\nu=0$, to
constrain the coefficients in expression~(\ref{eq:1}). It is worth
emphasizing that the EMRL results are not only exact, but also in
regimes that numerical relativity simulations cannot probe. More
specifically, we can exploit that in the EMRL the final spin cannot be
affected by 
the infinitesimally
small BH. In practice, this amounts to requiring that
\begin{equation}
a_{\rm fin}(a, \nu=0) = a\,,
\label{eq:2}
\end{equation}
which constrains four of the six coefficients
\begin{equation}
s_0 = s_2 = s_3 = 0\,, \qquad s_1=1\,.
\label{eq:3}
\end{equation}

\begin{figure}
\begin{center} 
\scalebox{0.4}{\includegraphics[angle=-90]{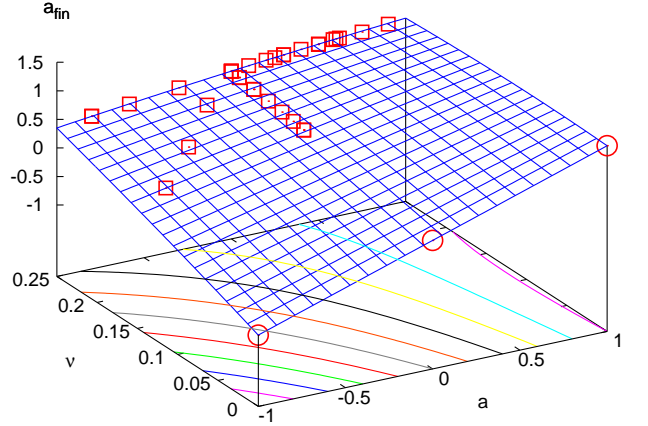}}
\caption{\label{plotone}Global dependence of the final spin on the
  symmetric mass ratio and on the initial spins as predicted by
  expression~(\ref{eq:4}). Squares refer to numerical estimates while
  circles to the EMRL constraints.}
\end{center}
\end{figure}

Additional but non-exact constraints 
can also be
applied by exploiting the knowledge, near the EMRL, of the functional
dependence of $a_{\rm fin}$ on the mass ratio. A convenient way of
doing this is suggested by BKL, and within this approach we 
perform a Taylor expansion of $a_{\rm fin}$ for $\nu \ll
1$ and determine that
\begin{eqnarray}
&&\left. a'_{\rm fin}\right|_{(a=1,\nu=0)} =
  2(\sqrt{3}/3 - 1 )\,, \quad
\left. a'_{\rm fin}\right|_{(a=0,\nu=0)}=
  2 \sqrt{3}\,, \nonumber \\
&&\left. a'_{\rm fin}\right|_{(a=-1,\nu=0)} =
 2 (1 + {19\sqrt{15}}/{45})\,,
\label{eq:5.1}
\end{eqnarray}
where $a'_{\rm fin} \equiv \partial a_{\rm fin}/\partial \nu$. The
coefficients in~(\ref{eq:1}) are then $s_4 = {\sqrt{3}}(19 \sqrt{5} -
75)/45$, $t_1 = 2 \sqrt{3}$, $t_0 = [{\sqrt{3}(15 - 19\sqrt{5}) -
90}]/{45}$. While this may seem a good idea, it leads to bad fits of
the data. We believe this is due to two distinct reasons: \textit{(i)}
the lack of accurate numerical data for near-extreme BHs,
\textit{i.e.}, $|a| \approx 1$, and which therefore leads to incorrect
estimates of the coefficients; \textit{(ii)}
expressions~(\ref{eq:5.1}) are analytic but not exact and should be
used with caution. There are, in fact, deviations from analyticity in
$\nu$ as $\nu \rightarrow 0$, and as revealed by the presence of
integer powers of $\nu^{1/5}$ during the transition between the last
stable orbit and the plunge (see~\citet{Buonanno_Damour:2000}). In the
case of non-spinning binaries ($a = 0$), it is now possible to verify
that the deviations are indeed very small~\citep{Damour_Nagar:2007a},
but this check is not possible for very large spins. In view of this
and to make the minimal number of assumptions, we retain the analytic
estimate only for the coefficient $t_1$, so that~(\ref{eq:1}) has five
out of ten coefficients constrained analytically
\begin{equation}
a_{\rm fin} = a + s_4 a^2\nu + s_5 a\nu^2 + t_0 a \nu +
              2\sqrt{3} \nu + t_2 \nu^2 + t_3 \nu^3\,.
\label{eq:4}
\end{equation}

Determining the remaining five coefficients from a least-square fit of
the available data yields
\begin{eqnarray}
s_4 = -0.129 \pm 0.012\,,& \quad 
s_5 = -0.384 \pm 0.261\,,& \nonumber\\
t_0 = -2.686 \pm 0.065\,,& \quad
t_2 = -3.454 \pm 0.132\,,& \nonumber\\
t_3 =  2.353 \pm 0.548\,,&&
\label{eq:4.1}
\end{eqnarray}
with surprisingly small residuals and large error-bars only for
$s_5$. The functional behaviour of expression~(\ref{eq:4}) and the
position of the numerical data points are shown in Fig.~\ref{plotone}.

\begin{figure}
\begin{center} 
\scalebox{0.4}{\includegraphics[angle=-0]{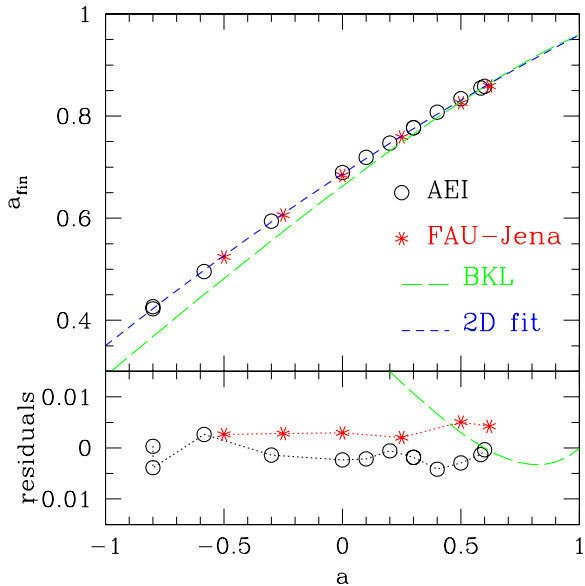}}
\caption{\label{plottwo}\textit{Upper panel:} Comparison of the
  numerical data with the 2D fit through~(\ref{eq:4}) in the case of
  equal-mass binaries, ($\nu = 1/4$). Empty circles
  indicate the AEI data~\citep{Rezzolla_etal:2007a}, stars the
  FAU-Jena data~\citep{Marronetti_etal:2007}], a long-dashed line the
  BKL, and a short-dashed one the fit. \textit{Lower panel:} residuals
  between the different estimates and the fit.}
\end{center}
\end{figure}

In the following we discuss the properties of the proposed fit,
providing evidence that it represents a very accurate description of
the available estimates, and discuss how to use it to make
astrophysically interesting predictions.

\textit{(i)} The estimate for the final spin in the case of equal
  masses and the comparison with available data and estimates is made
  in Fig.~\ref{plottwo}. The upper panel shows the numerical
  estimates, [circles for the AEI data~\citep{Rezzolla_etal:2007a} and
  stars for the FAU-Jena data~\citep{Marronetti_etal:2007}], the BKL
  estimate and our 2D fit through~(\ref{eq:4}). The lower panel shows
  the residuals between the different estimates and the 2D fit; these
  are always of a few percent only and become larger for the BKL
  estimate when $a \lesssim 0$.

\textit{(ii)} Despite the cubic dependence assumed in~(\ref{eq:1}),
  expression~(\ref{eq:4}) is only \textit{quadratic} with $a$. When
  $\nu=1/4$, it confirms what was obtained
  recently~\citep{Rezzolla_etal:2007a}, indicating that, for
  equal-mass binaries, the next order will be four.

\textit{(iii)} Using~(\ref{eq:4}) and~(\ref{eq:4.1}) we estimate that
  the minimum and maximum final spins for an equal-mass binary are
  $a_{\rm fin} = 0.3502 \pm 0.03$ and $a_{\rm fin} = 0.9590 \pm 0.03$,
  respectively.

\begin{figure}
\begin{center} 
\scalebox{0.4}{\includegraphics[angle=-0]{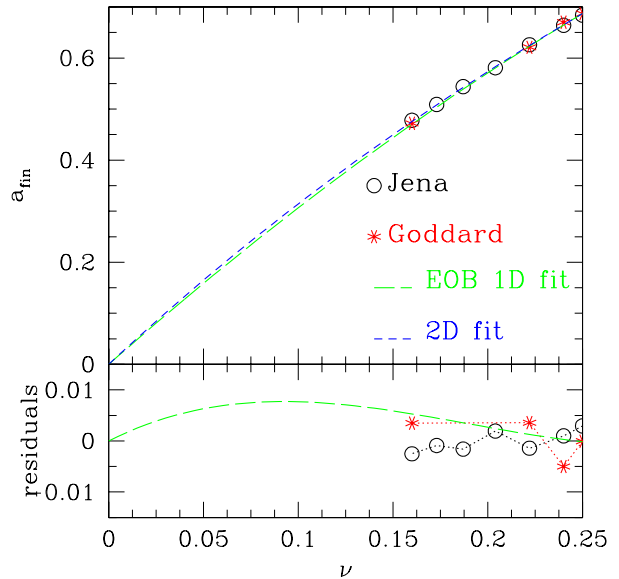}}
\caption{\label{plotthree}\textit{Upper panel:} Comparison of the
  numerical data with the 2D fit through~(\ref{eq:4}) in the case of
  nonspinning binaries. Empty circles indicate the Jena
  data~\citep{Berti_etal:2007}, stars the Goddard
  data~\citep{Buonanno_etal:2007a}], a long-dashed line the quadratic
  EOB fit~\citep{Damour_Nagar:2007a} and a short-dashed line our 2D
  fit. \textit{Lower panel:} residuals between the different estimates
  and the 2D fit.}
\end{center}
\end{figure}

\textit{(iv)} For nonspinning binaries,
  expression~(\ref{eq:4}) is cubic in $\nu$ and a comparison with the
  available data and the estimate from the EOB approach combined with
  test-mass limit predictions for the
  ringdown~\citep{Damour_Nagar:2007a} is shown in
  Fig.~\ref{plotthree}. In particular, the upper panel shows the
  numerical values, [empty circles for the Jena
  data~\citep{Berti_etal:2007} and stars for the Goddard
  data~\citep{Buonanno_etal:2007a}], a long-dashed line for the
  quadratic EOB 1D fit~\citep{Damour_Nagar:2007a} and a short-dashed
  line for our 2D fit. (Because it is very similar to the EOB
  estimate, we have not shown the BKL prediction.) The residuals in
  are shown in the lower panel. 

\begin{figure}
\begin{center} 
\scalebox{0.4}{\includegraphics[angle=-0]{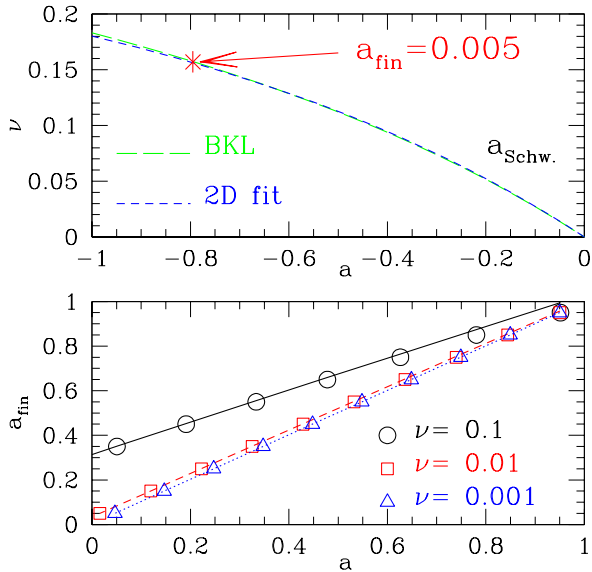}}
\caption{\label{plotfour} \textit{Upper panel:} Set of initial spins
  and mass ratios leading to a final Schwarzschild BH:
  \textit{i.e.}, $a_{\rm fin}(a,\nu)=0$. The two curves refer to
  the BKL estimate (long dashed) and to the 2D fit (short dashed),
  respectively. Indicated with a star is a numerical example leading
  to $a_{\rm fin} =0.005$. \textit{Lower panel:} Comparison between
  the BKL prediction (symbols) and the 2D fit (solid, dashed and
  long-dashed lines) near the EMRL. Different curves refer to
  different values of $\nu$ and the match is complete for $\nu=0$.}
\end{center}
\end{figure}

\textit{(v)} A physically useful condition that can be deduced from
  the 2D fit are the values of the initial spin and mass ratio that
  will lead to a final \textit{Schwarzschild}
  BH~\citep{Hughes_Blandford:2003,Buonanno_etal:2007b}. In practice
  this amounts to requiring $a_{\rm fin}(a,\nu)=0$ in~(\ref{eq:4}) and
  this curve in the $(a,\,\nu)$ plane is shown in the upper panel of
  Fig.~\ref{plotfour}. Binaries on the curve produce Schwarzschild
  BHs, while binaries above the curve start with a positive total
  angular momentum and end with a positive one; binaries below the
  curve, on the other hand, start with a positive total angular
  momentum and end with a negative one, \textit{i.e.}, with a global
  \textit{flip}. Also shown in the upper panel of Fig.~\ref{plotfour}
  is the prediction from BKL: $a_{_{\rm Schw.}}\vert_{_{\rm BKL}} =
  2\nu\sqrt{3}/(2\nu-1)$. The two estimates are very similar for all
  values of $\nu$ and small differences appear for $\nu \gtrsim 0.15$,
  where the BKL estimate is less accurate. Shown with a cross is the binary
  $p_1$ (\textit{cf.}, Table~\ref{tableone}) which yields a final BH
  with spin $a_{\rm fin}=0.005$. The numerical value is between the
  BKL prediction and the 2D fit.

\textit{(vi)} The BKL is expected to be particularly accurate for $\nu
  \ll 1$ and its prediction in this regime are captured very well by
  the 2D fit (of course the two predictions are identical for
  $\nu=0$). This is shown in the lower panel of Fig.~\ref{plotfour}
  with different curves referring to $\nu = 0.001\,,0.01$ and $0.1$;
  interestingly, the differences are small even for $\nu = 0.1$.

\begin{figure}[ht]
\begin{center} 
\vskip 0.25cm
\scalebox{0.4}{\includegraphics[angle=-0]{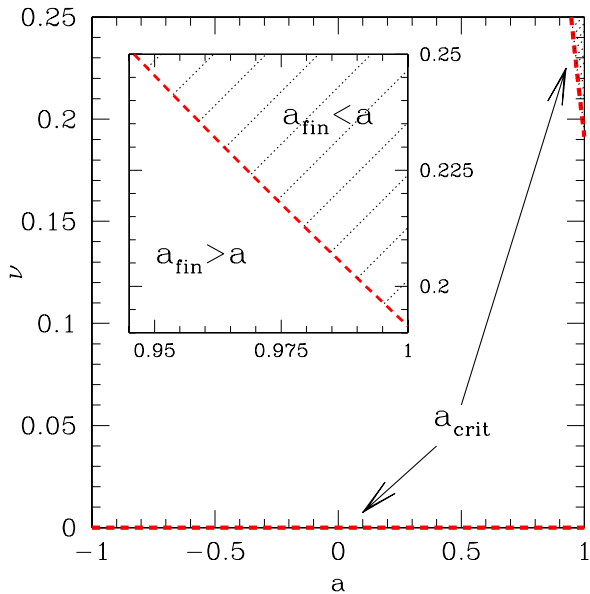}}
\caption{\label{plotfive} Critical values of the initial spin and mass
  ratio leading to a final BH having the same spin as the initial ones
  \textit{i.e.}, $a_{\rm fin}(a,\nu)=a$. A magnification is shown in
  the inset, where the dashed/non-dashed region refers to binaries
  \textit{spun-down/up} by the merger.}
\end{center}
\end{figure}

\textit{(vii)} It is simple to derive the value of $a$ which will
  produce a final BH with the \textit{same} spin as the initial
  ones. This amounts to requiring that $a_{\rm fin}(a,\nu)=a$
  in~(\ref{eq:4}) and the resulting solution is shown in
  Fig.~\ref{plotfive}; clearly, the axis $\nu=0$ is a trivial solution
  and a magnification of the behaviour away from the EMRL is shown in
  the inset. For equal-mass binaries the critical value is $a_{\rm
  crit} = 0.9460$, in very good agreement with the BKL estimate
  $a_{\rm crit} \gtrsim 0.948$~\citep{Buonanno_etal:2007b}. The
  minuteness of the region for which $a_{\rm fin} < a$ (dashed region)
  suggests that BHs from aligned-spins binaries are typically spun-up
  by mergers.

\textit{(viii)} It is easy to verify that by setting $\nu=1/4$ and $2
  a = a_1 +a_2$ in~(\ref{eq:4}), the coefficients $s_1$--$s_5$ and
  $t_0$--$t_3$ coincide, within the error-bars, with the coefficients
  $p_0$, $p_1$ and $p_2$ reported in~\citet{Rezzolla_etal:2007a} for
  equal-mass, unequal-spin binaries. The fact that the fit here is
  equivalent to, but has been independently derived from, the one for
  the equal-mass, unequal-spin binaries, is an indication of its
  robustness. Indeed, it is possible to extend~(\ref{eq:4}) to the
  whole $(a_1,a_2,\nu)$ space~\textit{i.e.}, to describe the final
  spin of generic aligned, unequal-spin, unequal-mass BH binaries, by
  replacing $a$ with $(a_1+a_2 q^2)/(1+q^2)$. The resulting expression
  reduces to~(\ref{eq:4}) for unequal-mass, equal-spin binaries, and
  to the one in~\citet{Rezzolla_etal:2007a} for equal-mass,
  unequal-spin binaries. Our suggested extension of~(\ref{eq:4}) to
  the $(a_1,a_2,\nu)$ space is the simplest one which recovers, for
  aligned spins, the well-tested limits of equal-mass, unequal-spins
  and unequal-mass, equal-spins. Work is in progress to validate this
  ansatz with numerical simulations.

A final comment is one of caution. The dependence of the final spin on
the mass ratio in the case of extreme aligned BHs is particularly
challenging to calculate and not yet investigated accurately by
numerical calculations. The predictions of expression~(\ref{eq:4}) in
this limit amount to mere extrapolations and are therefore accurate to
a few percent at most. As an example, when $a=1$, the fit~(\ref{eq:4})
is a non-monotonic function with maximum $a_{\rm fin} \simeq 1.029$
for $\nu \simeq 0.093$; this clearly is an artifact of the
extrapolation.

\section{Conclusions} 

Modelling the final spin in a generic binary BH merger is not trivial
given the large space of parameters on which this quantity depends.
We have shown that the results of recent simulations combined with
basic but exact considerations derived from the EMRL allow us to model
this quantity with a simple analytic expression in the case of BH
binaries having unequal masses and unequal spins which are aligned
with the orbital angular momentum. When compared with all other
estimates coming either from numerical calculations or from
approximation techniques, the estimates of the 2D fit show differences
which are of few percent at most.


\bigskip
\noindent We 
thank A. Buonanno, T. Damour,
S. A. Hughes, L. Lehner, A. Nagar, and B. S. Sathyaprakash for
discussions. We are grateful to D. Merritt for pointing out an error
in the interpretation of our results. The computations were performed on the
supercomputers at AEI, LITE, LSU, LONI and NCSA. Support comes also
through the DFG grant SFB/TR
~7.

\end{document}